\documentclass[12pt]{article}
\begin{document}

\begin{center}
{\Large  An algebraic method of classification of S-integrable discrete models}

\vskip 0.2cm

{Ismagil T. Habibullin}\footnote{e-mail: habibullinismagil@gmail.com}\\

{Ufa Institute of Mathematics, Russian Academy of Science,\\
Chernyshevskii Str., 112, Ufa, 450077, Russia}\\

\bigskip

{Elena V. Gudkova}\footnote{e-mail: elena.gudkova79@mail.ru}

{Department of Applied Mathematics and Mechanics,\\ Ufa State Petroleum Technical University,
 \\Kosmonavtov str., 1, Ufa, 450062, Russia }\\

\bigskip


\end{center}
\begin{abstract}

A method of classification of integrable equations on quad-graphs is discussed based on algebraic ideas. We assign a Lie ring to the equation and study the function describing the dimensions of linear spaces spanned by multiple commutators  of the ring generators. For the generic case this function grows exponentially. Examples show that for integrable equations it grows slower. We propose a classification scheme based on this observation.
\end{abstract}
\bigskip

{\it Keywords:} quad-graph equations, classification, characteristic vector fields, Lie ring, integrability
conditions, discrete KdV equation.

\section{Introduction}

Nowadays discrete models (quad--graph equations) of the form
\begin{equation}\label{ddhyp}
u_{1,1}=f(u,u_1,  \bar u_{1} )
\end{equation}
are intensively studied due to their applications in physics, geometry, biology etc. Let us explain the notations in (\ref{ddhyp}). Here the unknown is a function of two independent discrete variables $u=u(m,n)$. 
The subindices and bar stand for the shifts of the arguments: $u_k=u(m+k,n)$, $\bar{u}_k=u(m,n+k)$, $u_{1,1}=u(m+1,n+1)$.
Function $f$ is supposed to be locally analytic, it depends essentially on all three arguments. In other words equation (\ref{ddhyp}) can be rewritten in any of the following forms
\begin{eqnarray}
u_{1,-1}&=&f^{1,-1}(u,u_1, \bar u_{-1}),\label{1,-1}\\
u_{-1,1}&=&f^{-1,1}(u,u_{-1}, \bar u_{1}),\label{-1,1}\\
u_{-1,-1}&=&f^{-1,-1}(u,u_{-1}, \bar u_{-1}).\label{-1,-1}
\end{eqnarray}
At present various approaches are known for studying integrable discrete phenomena. The property of consistency around cube \cite{NijhoffWalker}, has been proposed as the integrability criterion for quadrilateral difference equations \cite{BobenkoSuris}, \cite{Nijhoff}, \cite{abs}. Symmetry approach to the classification of integrable systems is adopted to discrete case  \cite{LeviYamilov}, \cite{Xenitidis}, \cite{RasinHydon}, \cite{Mikhailov}, \cite{tongas}. Another characteristic property of an integrable equation is the vanishing of its algebraic entropy \cite{BellonViallet}. Alternative methods are used in \cite{NijhoffRamani}, \cite{GKP}, \cite{Hietarinta}.
In this article we suggest a new classification scheme for integrable discrete models.

About twenty years ago it was observed that characteristic Lie algebras, introduced in \cite{Shabat}, in case of integrable hyperbolic type PDE's like sine-Gordon and Zhiber-Shabat equations have a very specific property. The dimensions of the linear spaces spanned by multiple commutators of the generators grow essentially slower than in generic case \cite{ZhiberMukminov}. In \cite{ZhiberMur} the problem of rigorous formalization of the notion of "slow growth" has been discussed. A conjecture was suggested  and tested by applying to classify integrable equations of the form $u_{x,y}=f(u,u_x).$

Here we consider another formalization of this property of the characteristic vector fields and test it by taking as a touchstone a particular case of quad--graph equation (\ref{ddhyp}). 

The article is organized as follows. In section 2 we introduce characteristic vector-fields and define the test Lie ring. A connection between integrability and the test  ring is conjectured. In section 3 description of the test ring for discrete potential KdV equation is presented. The classification problem for the model of the form $u_{1,1}-u=g(u_1-\bar u_{-1})$ is investigated in section 4. The  result of classification is summarized in Theorem 5.

\section{Characteristic vector fields and classification scheme}

Let us define characteristic vector fields for the equation (\ref{ddhyp}). We begin with a very particular case when the equation admits an $n$-integral i.e. a function $I=I(u_{-j},u_{-j+1},...u_k)$ such that the equation holds $\bar DI=I$, where $\bar D$ is the shift operator: $\bar Dh(m,n)=h(m,n+1)$. This means that for any solution $u=u(m,n)$ of equation (\ref{ddhyp}) the value of the function $I$ does not depend on the variable $n$. In the coordinate representation condition $\bar DI=I$ means
\begin{equation}\label{I}
I(r_{-j+1},r_{-j+2},...,r,\bar u,f,f_1,...f_{k-1})=I(u_{-j},u_{-j+1},...u_k),
\end{equation}
where $r=f^{-1,1}(u,u_{-1},\bar u_1)$. Evidently the right hand side in (\ref{I}) does not depend on $\bar u_{1}$ hence the condition holds $\frac{\partial}{\partial \bar u_1}\bar D^{-1}I=0$ as well as $YI=0$ where $Y:=\bar D^{-1}\frac{\partial}{\partial \bar u_1}\bar D $. Due to the formula $\frac{\partial r}{\partial \bar u_1}=1/D^{-1}(\frac{\partial f}{\partial \bar u_1})$ where $D$ is the shift with respect to $m$, $Dh(m,n)=h(m+1,n),$  one finds (see also \cite{hab})
\begin{equation}\label{Y}
Y=\frac{\partial}{\partial u}+x\frac{\partial}{\partial  u_{1}}+\frac{1}{x_{-1}}\frac{\partial}{\partial  u_{-1}}+
xx_1\frac{\partial}{\partial  u_{2}}+\frac{1}{x_{-1}x_{-2}}\frac{\partial}{\partial u_{-2}}+ ...,
\end{equation}
where $x=\bar D^{-1}(\frac{\partial f(u,u_1,\bar u_1)}{\partial \bar u_{1}})=-\frac{\partial f^{1,-1}(u,u_1,\bar u_{-1})/\partial u}{\partial f^{1,-1}(u,u_1,\bar u_{-1})/\partial u_{1}}$. We call $Y$ characteristic vector field. Now turn back to the general case and define characteristic vector field $Y$ for chain (\ref{ddhyp}) as a formal series given by formula (\ref{Y}).

Denote through $T$ the set of vector fields obtained by taking all possible multiple commutators and linear combinations of the operators $X:=\frac{\partial}{\partial \bar u_{-1}}$ and $Y$ with coefficients depending on a finite number of the dynamical variables $\bar u_{-1}$, $u$, $u_{\pm 1}$, $u_{\pm 2}$, ... . Evidently the set $T$ has a structure of the Lie ring. We call it test ring of the equation (\ref{ddhyp}) in the direction of $n$. In a similar way one can define the test ring $\bar T$ in the direction of $m$.

Notice that for Darboux integrable equations of the form (\ref{ddhyp}) both rings $T$ and $\bar T$ are of finite dimension. Actually the test ring is a subset of the characteristic Lie ring \cite{hab}.

Denote through $V_j$ the linear space over the field of locally analytic functions  spanned by $X$, $Y$ and all multiple commutators of $X$ and $Y$ of order less than or equal to j such that:
$$V_0=\{X,Y\},\quad V_1=\{X,Y,[X,Y]\}, \quad \dots .$$

Introduce the function $\Delta(k)=\dim V_{k+1}-\dim V_k.$ The following conjecture is approved by numerous examples.

{\bf Conjecture (algebraic test)}. {\it Any integrable model of the form (\ref{ddhyp}) satisfies the following condition: there is a sequence of natural numbers $\left\{t_k\right\}_{k=1}^{\infty}$ such that $\Delta(t_k)\leq1$}.

Ring $T$ admits an automorphism, generated by the shift operator $D$, 
\begin{equation}\label{aut}
T\ni Z\stackrel{Aut}{\rightarrow} DZD^{-1}\in T,
\end{equation}
which plays crucial role in our further considerations. It is important that $X$ and $Y$ considered as operators on the set of functions depending on the variables $\bar u_{-1},u,u_{\pm1}, u_{\pm2},...$ satisfy the following conjugation relations
\begin{equation}\label{autXY}
DXD^{-1}=pX \quad\mbox{and}\quad DYD^{-1}=\frac{1}{x}Y,
\end{equation} 
where $p=D(\frac{\partial f^{-1,-1}(u,u_{-1},\bar u_{-1})}{\partial \bar u_{-1}})=\frac{1}{\partial f^{1,-1}(u,u_1,\bar u_{-1})/\partial \bar u_{-1}}$. Indeed, specify the coefficients of the operator 
$DXD^{-1}=\sum a_i\frac{\partial}{\partial u_i}+p\frac{\partial}{\partial \bar u_{-1}}$
by applying it to the dynamical variables and find that $a_i=DXD^{-1}u_i=0$ for any integer $i$. Moreover $p=DXD^{-1}\bar u_{-1}=DXf^{-1,-1}(u,u_{-1},\bar u_{-1})=D(\frac{\partial f^{-1,-1}(u,u_{-1},\bar u_{-1})}{\partial \bar u_{-1}})$. In a similar way one can prove the second formula. Really apply the operator $DYD^{-1}=\sum c_i\frac{\partial}{\partial u_i}+d\frac{\partial}{\partial \bar u_{-1}}$ to $u_i$
and find $c_j=D(Yu_{j-1})=Yu_j.$ Then evaluate $d=DYD^{-1}\bar u_{-1}=f^{-1,-1}_u+\frac{1}{x_{-1}}f^{-1,-1}_{u_{-1}}$. Since $u_{-1,-1}=f^{-1,-1}(u,u_{-1},\bar u_{-1})$ one gets the equation $u=f^{-1,-1}(f(u,u_1,\bar u_1),u_1,\bar u_1)$. Let us differentiate  it with respect to $\bar u_{1}$ and find
$D\bar D(\frac{\partial f^{-1,-1}}{\partial u})\frac{\partial f}{\partial \bar u_{1}}+D\bar D(\frac{\partial f^{-1,-1}}{\partial u_{-1}})=0$ or the same $\frac{\partial f^{-1,-1}}{\partial u}+ \frac{1}{D^{-1}\bar D^{-1}(\frac{\partial f}{\partial \bar u_{1}})}\frac{\partial f^{-1,-1}}{\partial u_{-1}}=0$. Now due to the equation $x=\bar D^{-1}(\frac{\partial f}{\partial \bar u_{1}})$
 one concludes that $d=0$.

{\bf Lemma} 1. Suppose that $Z=\sum_{-\infty}^{\infty} b_j\frac{\partial}{\partial  u_{j}}\in T$ satisfies the following two conditions: 1) $DZD^{-1}=cZ$ for some function $c$ and 2) $b_{j_0}\equiv 0$ for some fixed value of $j=j_0$.  Then $T=0.$

{\bf Proof}. We have $DZD^{-1}=\sum_{-\infty}^{\infty} D(b_{j-1})\frac{\partial}{\partial  u_{j}}=\sum_{-\infty}^{\infty} cb_j\frac{\partial}{\partial  u_{j}}$, if $c=0$ then $\forall j \quad D(b_{j-1})=0 $  and lemma is proved. If $c\neq0$ then one finds $b_{j_0-k}=cD^{-k}b_{j_0}$ and $b_{j_0+k}=D^{k}b_{j_0}/c$ for $k>0$ which completes the proof.

\section{Discrete potential Korteweg--de Vries equation}

In this section we give a complete description of the test Lie ring $T$ for the discrete potential KdV equation \cite{NijhoffCapel}
\begin{equation}\label{dkdv}
u_{1,1}=u+\frac{1}{u_1-\bar u_1},
\end{equation}
which is a very well known example of integrable model of the form (\ref{ddhyp}). For the invariance of the equation with respect to the change $m\leftrightarrow n$ the rings $T$ and $\bar T$ are isomorphic. 

For this equation the factors $x$ and $p$ in the formula (\ref{autXY}) are the same $p=x=(u_1-\bar u_{-1})^2.$ 
Define a sequence of the vector fields $R_1=[X,Y],$ $P_1=[X,R_1],$ $Q_1=[Y,R_1]$ and then
$R_{k+1}=[X,Q_k],$ $P_k=[X,R_k],$ $Q_k=[Y,R_k]$, $k\geq1.$

{\bf Theorem} 1. The sequence $X,Y,R_1,P_1,Q_1,R_2,P_2,Q_2,...$ constitutes a basis of the characteristic ring $T$ of equation (\ref{dkdv}). 

{\bf Proof}. It is easily checked that $Xx=-2\sqrt x$, $Yx=2x\sqrt x$, $Xy=-2y\sqrt y$, $Yy=2\sqrt y$, where $y:=D^{-1}x=x_{-1}$. By using the relations $DXD^{-1}=xX$ and $D(yY)D^{-1}=Y$,  one can find the following equations
\begin{eqnarray}
D(R_1-2\sqrt yY)D^{-1}&=&R_1-2\sqrt xX,\label{R1}\\
D(P_1-2\sqrt yR_1+2yY)D^{-1}&=&x(P_1+2X),\label{P1}\\
D(y(Q_1-2Y))D^{-1}&=&Q_1+2\sqrt xR_1-2xX.\label{Q1}
\end{eqnarray}
A similar formula for $R_2$ is as follows
\begin{eqnarray}
D(R_2-2\sqrt yQ_1)D^{-1}&=&R_2+2\sqrt xP_1.\label{R2}
\end{eqnarray}
For $P_2$ and $Q_2$ we have
\begin{eqnarray}
D(P_2+2\sqrt yR_2-2yQ_{1})D^{-1}&=&x(P_2-2P_1),\label{P2}\\
D(y(Q_2-2Q_1))D^{-1}&=&Q_2+2\sqrt xR_2+2xP_1.\label{Q2}
\end{eqnarray}

One can prove by induction that for any $j>1$
\begin{eqnarray}
D(R_j-2\sqrt yQ_{j-1})D^{-1}&=&R_j+2\sqrt xP_{j-1},\label{Rj}\\
D(P_j+2(-1)^j\sqrt yR_j+2(-1)^{j-1}yQ_{j-1})D^{-1}&=&x(P_j-2P_{j-1}),\label{Pj}\\
D(y(Q_{j}-2Q_{j-1}))D^{-1}&=&Q_{j}+2\sqrt xR_j-2xP_{j-1}X,\label{Qj}
\end{eqnarray}
and then $[X,P_j]=0,$ $[Y,Q_j]=0,$ $[Y,P_j]=[X,Q_j],$ $[R_j,P_k]=P_{k+j},$ $[R_j,Q_k]=-Q_{k+j},$ $[R_j,R_k]=0,$ $[P_j,Q_k]=-R_{k+j+1},$ $[P_j,P_k]=0,$ $[Q_j,Q_k]=0.$

Give a matrix representation of the Lie algebra generated by the same operators $X$, $Y$.
Let
$$
X\rightarrow\lambda\sigma_+,\quad Y\rightarrow\lambda\sigma_-,\quad R_1\rightarrow\lambda^2\sigma_3,$$
then 
$$
P_k\rightarrow-2^{k}\lambda^{2k+1}\sigma_+,\quad Q_k\rightarrow2^{k}\lambda^{2k+1}\sigma_-,\quad R_k\rightarrow2^{k-1}\lambda^{2k}\sigma_3,$$
where
\begin{equation}\label{sigma}
\sigma_+=\left(%
\begin{array}{cc}
  0 & 1  \\
  0 & 0 \\
\end{array}%
\right), \quad \sigma_-=\left(%
\begin{array}{cc}
  0 & 0 \\
  1 & 0 \\
\end{array}%
\right), \quad
\sigma_3=\left(%
\begin{array}{cc}
  1& 0  \\
  0 & -1 \\
\end{array}%
\right), \end{equation}

It can easily be checked that for equation (\ref{dkdv}) function $\Delta=\Delta(k)$ is periodic with period 2: $\Delta(2k)=1$, $\Delta(2k+1)=2.$

\section{Equations of the form $u_{1,1}=u+g({u_1-\bar u_1})$}

Apply the {\bf Conjecture} to the following particular class of discrete model (\ref{ddhyp})
\begin{equation}\label{pcase}
u_{1,1}=u+g({u_1-\bar u_1}).
\end{equation}
Here $g$ is a function to be determined. 

{\bf Classification scheme}. Study separately the following four different cases of equation (\ref{pcase}):
\begin{enumerate}
\item[i)] $\Delta(0)<\Delta_{max}(0)=1$;
\item[ii)] $\Delta(0)=\Delta_{max}(0)$, $\Delta(1)<\Delta_{max}(1)=2$; 
\item[iii)] $\Delta(0)=\Delta_{max}(0)$, $\Delta(1)=\Delta_{max}(1)$, $\Delta(2)<\Delta_{max}(2)=3$;
\item[iv)] $\Delta(0)=\Delta_{max}(0)$, $\Delta(1)=\Delta_{max}(1)$, $\Delta(2)=\Delta_{max}(2)$ and  $\Delta(k)\leq1$ for some $k>2$.
\end{enumerate}
Where $\Delta_{max}(k)$ stands for the greatest value of $\Delta(k)$ for equation (\ref{ddhyp}) when $f(u,u_1,\bar u_{1})$ ranges the class of arbitrary functions.

It is surprising that for equation (\ref{pcase}) investigation of the first three particular cases $i)$-$iii)$ allows one to extract a very short list of equations which are expected to be integrable. This list is exhaustive because the case $iv)$ is never realized (see below Corollaries of Theorems 2 and 3).

Introduce vector fields $R_1=[X,Y]$, $P_1=[X,R_1],$ $Q_1=[Y,R_1]$, $R_2=[X,Q_1]$, $W=[Y,Q_1]$, $Z=[X,P_1]$. Using these vector fields we can span in addition to $V_0$ and $V_1$ two more linear spaces:
$$V_2=V_1+\{P_1,Q_1\},\quad V_3=V_2+\{W,Z,R_2\}.$$

In order to evaluate $\Delta (k)$ we will use the automorphism (\ref{aut}). At first specify the factors $x$ and $p$ in formula (\ref{autXY}) for the case (\ref{pcase}): $x=p=-g'(g^{-1}(u_1-\bar u_{-1}))$, where the function  $\beta=g^{-1}(\alpha)$ is the inverse to the function $\alpha=g(\beta)$. Inversely, knowing $x=x(u_1-\bar u_{-1}))$ one can recover $g(\beta)$ by using the equation
\begin{equation}\label{restoreg}
\beta=g^{-1}(\alpha)=\int (g^{-1}(\alpha))'d\alpha=\int\frac{d\alpha}{g'(g^{-1}(\alpha)}=-\int\frac{d\alpha}{x(\alpha)}.
\end{equation}
Let us specify the action of the operators $X$ and $Y$ on the variable $x$. Evidently, $Xx=-x'$, $Yx=xx'$. It is found by direct calculation that
\begin{eqnarray}
DR_1D^{-1}&=&R_1+\frac{x'}{x}Y-x'X,\nonumber\\
DP_1D^{-1}&=&xP_1+x'R_1-rY+xrX,\quad r=x^{''}-\frac{x^{'2}}{x},\nonumber\\
DQ_1D^{-1}&=&\frac{1}{x}Q_1+\frac{x'}{x}R_1+\frac{x^{''}}{x}Y-x^{''}X,\nonumber\\
DWD^{-1}&=&\frac{1}{x^2}W+(\frac{2x^{''}}{x}-\frac{x^{'2}}{x^2})R_1+\frac{x^{'''}}{x}Y-x^{'''}X,\label{tableconjec}\\
DZD^{-1}&=&x^2Z+(x^{'2}-2xx^{''})R_1+qY-xqX,\quad q=xx^{'''}-2x^{'}x^{''}+\frac{x^{'3}}{x},\nonumber\\
DR_2D^{-1}&=&R_2+\frac{x'}{x}Q_1+x'P_1+\frac{x^{'2}}{x}R_1-sY+xsX,\quad s=x^{'''}-\frac{x'x^{''}}{x},\nonumber
\end{eqnarray}

Study the set $G$ of all multiple commutators of $X$ and $Y$.

{\bf Lemma 2}. The coefficients of any operator in $G$ are functions of a finite number of the variables $x$, $x_{\pm1}$, $x_{\pm2},...\,.$

{\bf Proof}. Since $x=x(u_1-\bar u_{-1})$ then evidently one can write $x'=\phi(x)$ for some function $\phi$. Then $X(x)=-\phi(x)$ and 
$Y(x)=x\phi(x)=:\psi(x)$. By using the conjugation relations $DXD^{-1}=xX$, $DYD^{-1}=\frac{1}{x}Y$ one derives that $X(x_j)=\phi^j(x,x_1,...x_j)$ and $Y(x_j)=\psi^j(x,x_1,...x_j)$. Similarly $X(x_{-j})=\phi^{-j}(x,x_{-1},...x_{-j})$ and $Y(x_{-j})=\psi^{-j}(x,x_{-1},...x_{-j})$. Now evidently $R_1=X(x)\frac{\partial}{\partial u_1}+X(\frac{1}{x_{-1}})\frac{\partial}{\partial u_{-1}}+X(xx_1)\frac{\partial}{\partial u_2}+\cdots$ satisfies the statement of the lemma. Due to the formulas $R_1(x)=X(x)x'$ and $DR_1D^{-1}=R_1+\frac{x'}{x}Y-x'X$ one gets $R_1(x_j)=\phi^j(x,x_1,...x_j)$. Obviously proof of the Lemma 2 can be completed by using induction.

{\bf Theorem 2}. If a chain of the form (\ref{pcase}) satisfies one of the conditions $i)-iii)$ of the {\bf Classification scheme} then function $x=x(\alpha)$ solves the following ordinary differential equation
\begin{equation}\label{ode}
x^{'2}=(x^2+1)\gamma+x\nu
\end{equation}
with $\gamma, \nu$ being constant.

{\bf Proof of the theorem 2}. Begin with the case $i)$, suppose that $\Delta (0)=0$, then we have $R_1=\lambda X +\mu Y$. It is evident that $R_1=X(x)\frac{\partial}{\partial  u_{1}}+...$ and hence $\lambda=\mu=0$, therefore $R_1=0$.  By applying the automorphism above to both sides of the last equation one finds 
$$\frac{x'}{x}Y-x'X=0.$$
Since $X$ and $Y$ are linearly independent we get equation $x'=0$ which is a particular case of (\ref{ode}). Evidently its solution is $x=c$ and due to (\ref{restoreg}) it can be found that $\beta=g^{-1}(\alpha)=-\frac{1}{c}\alpha+c_1$. Thus our equation $\alpha=g(\beta)$ (see (\ref{pcase})) is linear $u_{1,1}-u=-c(u_1-\bar u_1+c_1)$. For this case $\dim T=2$ so that $\Delta (k)=0$ for $k\geq0$. In a similar way one checks that condition $ii)$ leads to (\ref{ode}). Indeed  suppose that $\Delta(0)=1$ and $\Delta(1)<2$ then we have 
\begin{equation}\label{Delta1}
P_1=\nu Q_1+\epsilon R_1.
\end{equation}
Here due to Lemma 2 functions $\nu$ and $\epsilon$ might depend only on $x$, $x_{\pm 1}$, $x_{\pm 2}\dots$.
Apply the automorphism (\ref{aut}) to both sides of (\ref{Delta1}) then simplify due to equations (\ref{tableconjec}):
$$x(\nu Q_1+\epsilon R_1)+ x'R_1-rY+xrX=D(\nu)(\frac{1}{x}Q_1+\frac{x'}{x}R_1+\frac{x^{''}}{x}Y-x^{''}X)+ D(\epsilon)(R_1+\frac{x'}{x}Y-x'X).$$
Comparison of the coefficients before linearly independent operators gives rise to the conditions 
\begin{eqnarray}
Q_1:&&\quad x\nu=\frac{1}{x}D(\nu); \nonumber\\
R_1:&&\quad x'+ x\epsilon=\frac{x'}{x}D(\nu)+D(\epsilon); \nonumber\\
Y:&&\quad -r=\frac{x'}{x}D(\epsilon)+\frac{x''}{x}D(\nu); \nonumber\\
X:&&\quad xr=-x'D(\epsilon)-x''D(\nu). \nonumber
\end{eqnarray}
By analyzing these equalities one concludes that equation (\ref{Delta1}) holds if and only if the following three conditions 
valid: $\nu=0$, $\epsilon=const,$ $x'=\epsilon(1-x).$ Indeed under these conditions the last two equations above are satisfied automatically. Similarly one can check that $Q_1=\nu P_1-\epsilon R_1$ is equivalent to the same three conditions. Hence if $\Delta (1)<2$ then $\Delta (1)=0$ and therefore $\Delta (k)=0$ for any natural $k\geq1$. In this case $\dim T=3$.

Suppose now that $\Delta (0)=1$, $\Delta (1)=2$ and $\Delta (2)\leq2$ which corresponds to the case $iii)$. Start with the case when $Z$ is linearly expressed through the other vector fields in $V_3$:
\begin{equation}\label{z}
Z=\alpha X+\beta Y+\gamma R_1+ \delta P_1+\epsilon Q_1+\phi R_2+\psi W.
\end{equation}
Evidently $\alpha=\beta=0$, since $X=\frac{\partial}{\partial \bar u_{-1}}$ and $Y=\frac{\partial}{\partial  u}+x\frac{\partial}{\partial  u_{1}}+...$ and $Z$ does not contain the terms  $\frac{\partial}{\partial \bar u_{-1}}$ and 
$\frac{\partial}{\partial u}$.

By applying the automorphism to both sides of (\ref{z}) and comparing the coefficients before linearly independent operators we find
\begin{eqnarray}
W:&\,x^2\psi&=D(\psi)\frac{1}{x^2},\nonumber \\
R_2:&\,x^2\phi&=D(\phi),\nonumber \\
Q_1:&\,x^2\epsilon &=D(\epsilon)\frac{1}{x}+\frac{x'}{x}\phi,\nonumber \\
P_1:&\,x^2\delta &=D(\delta)x+x'\phi,\nonumber \\
R_1:&\, x^2\gamma &+x^{'2}-2xx^{''}=D(\delta)x'+D(\gamma).\nonumber 
\end{eqnarray}
Since $x=x(u_1-\bar u_{-1})$ we have $\psi=0$, $\phi=0$, $\epsilon=0$, $\delta=0$, $\gamma=const$. Comparison of the coefficients of $X$ and $Y$ gives one more equation $xq=\gamma x'$. Finally we get two ordinary differential equations for $x$:
$$x^2x^{'''}-2xx'x^{''}+x^{'3}=\gamma x', \quad (x^2-1)\gamma+x^{'2}-2xx^{''}=0. $$
The consistency condition of these equations is equivalent to equation (\ref{ode}). In this case $Z=\gamma R_1$. It is remarkable that  $x$ solves equation (\ref{ode}) if and only if $W$ is linearly expressed through the other elelments of $V_3$ and then $W=\gamma R_1$. And the last possibility is when $R_2$ is linearly expressed through $X$, $Y$, $R_1$, $P_1$, $Q_1$, $W$, $Z$. In this case $x$ solves the equation $x'=0$. The proof of the theorem is completed. 

In order to find $x=x(\alpha)$ evaluate the integral
\begin{equation}\label{x}
H(x):=\int\frac{dx}{\sqrt{(x^2+1)\gamma+x\nu}}=\alpha-\alpha_0. 
\end{equation}
The answer is given by the formula
\begin{equation}\label{H}
H(x)=\frac{1}{\sqrt{\gamma}}\ln(2\sqrt{(x^2+1)\gamma+x\nu}+2x+b).
\end{equation}
Now find $x$ by solving the equatin $H(x)=\alpha-\alpha_0$. 
$$x(\alpha)=\frac{1}{4}e^{\sqrt{\gamma}(\alpha-\alpha_0)}-\frac{\nu}{2\gamma}-(1-\frac{\nu^2}{4\gamma^2})e^{-\sqrt{\gamma}(\alpha-\alpha_0)}.$$
In order to get the corresponding quad-graph equation (\ref{pcase}) integrate again:
\begin{equation}\label{g}\beta=g^{-1}(\alpha)=-\int\frac{d\alpha}{x(\alpha)}.\end{equation}
Then the equation searched is given by the formula (\ref{pcase}). Our conjecture is that equation (\ref{pcase}) with $g$ found from (\ref{x})-(\ref{g}) is S-integrable. Calculating the integral (\ref{g}) we find the list of equations desired.  
 
{\bf Corollary of Theorem 2}. If for chain (\ref{pcase}) one of the conditions $i)-iii)$ is satisfied then the chain is of one of the form:
\begin{eqnarray}
&i)\,&u_{1,1}-u=c(u_1-\bar u_1-\beta_0)\quad \mbox{for} \quad \gamma=\nu=0,\nonumber\\
&ii)\,&ae^{\sqrt{\gamma}(u_{1,1}-u)}=b+ce^{-\sqrt{\gamma}(u_1-\bar u_1)}\quad \mbox{for}\quad \nu=-2\gamma,\nonumber\\
&iii1)\,&(u_{1,1}-u-\alpha_0)(u_1-\bar u_1-\beta_0)=\frac{1}{4\nu}\quad \mbox{for}\quad \gamma=0, \nu\neq0\nonumber\\
&iii2)\,&ae^{\sqrt{\gamma}(u_{1,1}-u)}=b+ce^{\sqrt{\gamma}(u_1-\bar u_1)}\quad \mbox{for}\quad \nu=2\gamma,\nonumber\\
&iii3)\,&(4-\frac{\nu^2}{\gamma^2})e^{\sqrt{\gamma}(u_{1,1}-u-\alpha_0)}+\frac{\nu}{\gamma}=2\tanh{\frac{\sqrt{\gamma}(u_1-\bar u_1)-\beta_0}{2}}\quad \mbox{for}\quad \gamma\neq0, \quad \nu\neq\pm2\gamma,\nonumber
\end{eqnarray}
where $a$, $b$, $c$, $\alpha_0$, $\beta_0$ are constants. 
The third equation is nothing else but the well-known dpkdv equation (\ref{dkdv}). Equations $ii)$ and $iii2)$ are reduced to one another by a simple change of independent variables $m\leftrightarrow n$. 
By changing the variables $u=\frac{1}{\sqrt{\gamma}}\ln v$ we reduce the last two equations to a bilinear form 
\begin{eqnarray}
&iii2')\,&a\bar v_1v_{1,1}=bv\bar v_1+cv v_1,\nonumber\\
&iii3')\,&(\alpha_1v_1+\alpha_2\bar v_1)v_{1,1}+v(\alpha_3v_1+\alpha_4\bar v_1)=0,\nonumber
\end{eqnarray}
where $\alpha_1=(2-\frac{\nu^2}{2\gamma^2})e^{-\sqrt{\gamma}\alpha_0-\beta_0}\neq0,$ $\alpha_2=(2-\frac{\nu^2}{2\gamma^2})e^{-\sqrt{\gamma}\alpha_0}\neq0,$ $\alpha_3=(\frac{\nu}{2\gamma}-1)e^{-\beta_0}\neq0,$ $\alpha_4=(\frac{\nu}{2\gamma}+1)\neq0.$ Equation $iii2')$ was introduced in \cite{NijhoffRamani}, recently it was proved that this equation admits infinite series of higher symmetries \cite{Rasin}. Equation  $iii3')$ is a particular case of the Hietarinta-Viallet  equation \cite{HietarintaViallet}:
\begin{equation}\label{VialletHietarinta}
c_1u\bar{u}_1+c_2u_1u_{1,1}+c_3uu_1+c_4\bar{u}_1u_{1,1}+c_5(uu_{1,1}+u_1\bar{u}_1) = 0.
\end{equation}
with $c_5=0$. Recently it was proved that equation (\ref{VialletHietarinta}) passes the symmetry test \cite{YL}. 

Study the set $G$ consisting of $X$, $Y$ and their all multiple commutators. Define the order of an element $Z\in G$ as a number of its factors $X$ and $Y$ minus one. For instance $ord [X,Y]=1$,  $ord [X,[X,Y]]=2$ and so on. Define the degree $deg (Z)$ of $Z$ as the exponent $k$ in the conjugation relation $DZD^{-1}=x^kZ+...$, where the tail is a linear combination of the elements with orders less than $ord(Z)$. Denote through $G_{i,j}$ a subset of $G$ containing elements with the order $i$ and the degree $j$. Let $G_i=\bigcup_{j}G_{i,j}$ be the union of $G_{i,j}$. Evidently the set $G_{i,i-1}$ (as well as $G_{i,-i+1}$) contains the only element $Z_{i,i-1}=ad_X^i(Y)$ up to the factor $-1$ (correspondingly, the only element $Z_{i,-i+1}=ad_Y^i(X)$ up to the factor $-1$). Here the operator $ad$ is defined as follows $ad_X(Y)=[X,Y]$. 

{\bf Theorem} 3. Suppose that $Z_{k,k-1}$ (or $Z_{k,-k+1}$) is in the basis of the linear space $V_k\supset G_k$ for any $k$: $3\leq k<N$, but $Z_{N,N-1}$ (respectively $Z_{N,-N+1}$) is linearly expressed through the other operators in $V_N$ then function $x=x(u_1-\bar u_{-1})$ solves an equation of the form $x'=\epsilon(x-1)$ with a constant coefficient $\epsilon$.

{\bf Lemma} 3. For any $k\geq 3$ we have
\begin{equation}\label{zconjugation}
DZ_{k+1,k}D^{-1}=x^kZ_{k+1,k}-c_kx'x^{k-1}Z_{k,k-1}+\cdots\, ,
\end{equation}
where $c_k>0$ and the tail contains a linear combination of the operators with the order less than $k$. 

Prove the Lemma 3 by method of mathematical induction.  From the list of equations 
(\ref{tableconjec}) one gets for $Z=Z_{3,2}$ and $R_1=Z_{1,0}$ the following representation
\begin{equation}\label{z32}
DZ_{3,2}D^{-1}=x^2Z_{3,2}+(x^{'2}-2xx^{''})Z_{1,0}+qY-xqX,\quad q=xx^{'''}-2x^{'}x^{''}+\frac{x^{'3}}{x}
\end{equation}
showing that the statement is true for the case $k=3$. Suppose now that $DZ_{k,k-1}D^{-1}=x^{k-1}Z_{k,k-1}-c_{k-1}x'x^{k-2}Z_{k-1,k-2}+\cdots\,$ and evaluate $DZ_{k+1,k}D^{-1}$:
$$DZ_{k+1,k}D^{-1}=[xX,x^{k-1}Z_{k,k-1}-c_{k-1}x'x^{k-2}Z_{k-1,k-2}+\cdots\,]=x^kZ_{k+1,k}-c_kx'x^{k-1}Z_{k,k-1}+\cdots\,$$
where $c_k=c_{k-1}+(k-1)>c_{k-1}>0$. The proof is completed.

{\bf Proof of the Theorem 3}. Suppose that 
\begin{equation}\label{decompositionzN}
Z_{N,N-1}=\sum_{ord (Z_\nu)=N}a_{\nu}Z_{\nu}+\sum_{ord (Z_\mu)=N-1}b_{\mu}Z_{\mu}+\cdots\, ,
\end{equation}
where $Z_\nu$ and $Z_\mu$ range the basis of $V_N$ and the tail contains a linear combination of the operators of less order.
Apply the automorphism $D\cdot D^{-1}$ (see (\ref{aut}) above) to both sides of (\ref{decompositionzN}): 
\begin{eqnarray}
&&x^{N-1}(\sum_{ord (Z_\nu)=N}a_{\nu}Z_{\nu}+\sum_{ord (Z_\mu)=N-1}b_{\mu}Z_{\mu}+\cdots)-x^{N-2}x'c_{N-1}Z_{N-1,N-2}+\cdots=\nonumber\\
&&=\sum_{ord (Z_\nu)=N}D(a_{\nu})(x^{k_{\nu}}Z_{\nu}+\cdots) +\sum_{ord (Z_\mu)=N-1}D(b_{\mu})(x^{k_{\mu}}Z_{\mu}+\cdots). \nonumber
\end{eqnarray}
Compare the coefficients before $Z_{\nu}$ and get 
\begin{equation}\label{comparison}
x^{N-1}a_{\nu}=x^{k_{\nu}}D(a_{\nu}), \quad k_{\nu}\neq N-1.
\end{equation}
Due to Lemma 2 functions $a_{\nu}$ and $b_{\mu}$ depend on $x$ and its shifts. Moreover it follows from (\ref{comparison}) that $a_{\nu}$ cannot depend on $x$, $x_{\pm 1}$, $x_{\pm 2},\dots$ at all. Therefore the only possibility is $a_{\nu}=0$. Compare now the coefficients before $Z_{N-1,N-2}$ and find:
\begin{equation}\label{comparison2}
x^{N-1}b-c_{N-1}x^{N-2}x'=x^{N-2}D(b), 
\end{equation}
where $b$ is the coefficient of $Z_{N-1,N-2}$ in the expansion (\ref{decompositionzN}). A simple analysis of equation (\ref{comparison2}) shows that $b$ is constant. Thus (\ref{comparison2}) is equivalent to the equation $x'=\epsilon(x-1)$ with $\epsilon=b/c_{N-1}$.

{\bf Corollary of Theorem 3}. The case $iv)$ of the {\bf Classification scheme} is never realized.

{\bf Proof}. Suppose in contrary that such a case is realized. Then at least one of the vector fields $Z_{k,k-1}$ or $Z_{k,-k+1}$ should be linearly expressed through other elements of $V_k$, otherwise $\Delta(k)\geq2$. Therefore due to the Theorem 3 we have $x'=\epsilon(x-1)$ which corresponds to the case $ii)$ $\Delta(0)=1$, $\Delta(1)<2$. The contradiction shows that our assumption is not true. The proof is completed.

Study in details test rings $T$ and $\bar T$ for equation $iii1)$ from the list in {\bf Corollary of Theorem 2}. It was mentioned earlier that the equation is reduced to $ii)$ just by changing the independent variables $m\leftrightarrow n$. As it was shown above for $ii)$ dimension of the ring $T$ equals 3. This implies immediately that for the equation $iii2)$ $\dim \bar T=3$. Now concentrate on the ring $T$ for this equation. Without loss of generality we take $\gamma=1$ and get
\begin{equation}\label{iii2}
ae^{u_{1,1}-u}=b+ce^{u_1-\bar u_1}.
\end{equation}
We have already proved that for (\ref{iii2}) function $x=x(\alpha)$ solves the equation $x'=x+1.$ The characteristic vector fields $X$ and $Y$ act on the variables $x$, $y=D^{-1}x$ as follows: $Xx=-x'=-x-1,$ $Xy=-yy'=-y^2-y,$ $Yx=xx'=x^2+x,$ $Yy=y+1.$

{\bf Theorem} 4. The sequence $X,Y,R_1,P_1,Q_1,R_2,P_2,Q_2,...$ defined as follows $R_1=[X,Y],$ $P_1=[X,R_1],$ $Q_1=[Y,R_1]$ and then
$R_{k+1}=[X,Q_k],$ $P_k=[X,R_k],$ $Q_k=[Y,R_k]$, $k\geq1$ constitutes a basis of the test ring $T$ of equation (\ref{iii2}). 

{\bf Scheme of the proof}. Begin with the conjugation relations for multiple commutators. For instance, $D(R_1-Y+X)D^{-1}=R_1+Y-X.$ It can easily be proved by induction that for $k\geq1$
\begin{eqnarray}
D(P_k-R_k)D^{-1}&=&x(P_k+R_k),\nonumber\\
D(Q_k-R_k)D^{-1}&=&\frac{1}{x}(Q_k+R_k),\nonumber\\
D(R_{k+1}-Q_k-P_k+R_k)D^{-1}&=&R_{k+1}+Q_k+P_k+R_k.\nonumber
\end{eqnarray}
These relations imply that $[Y,P_k]=[X,Q_k]$ and $[X,P_1]=[Y,Q_1]=R_1.$ Then one obtains that 
\begin{eqnarray}
 &[R_k,R_j]&=[Q_k,Q_j]=[P_k,P_j]=0, \nonumber\\
 &[R_k,P_j]&=P_{k+j}-Q_{k+j-1}, \quad [R_k,Q_j]=-Q_{k+j}+P_{k+j-1},\nonumber\\
 &[P_k,Q_j]&=-R_{k+j+1}+R_{k+j-1}.\nonumber
\end{eqnarray}
Now one concludes that for equation (\ref{iii2}) test ring $T$ satisfies the conditions $\Delta(2k)=1$ and $\Delta(2k+1)=2$ as well as for the dpkdv equation $iii1)$, see {\bf Example} above. Thus this equation passes completely our algebraic test. For the fifth case $iii3)$ in the list of Corollary of Theorem 2 the test rings are not yet studied in detail.

Let us summarize the results above in the following theorem.

{\bf Theorem} 5. Suppose that an equation of the form (\ref{pcase}) passes the {\bf algebraic test} (see {\bf Conjecture}) then it is of one of the form given in {\bf Corollary of Theorem 2}.

\section{Conclusions.}

A new classification scheme for quad-graph equations is suggested based on distributions of characteristic vector fields (see, for instance \cite{DoubrovZelenko}). An exhaustive list of equations of the form 
$u_{1,1}-u=g(u_1-\bar u_{-1})$
passing the suggested integrability test is obtained. It is important that all the equations found are known to be integrable. The {\bf Conjecture} is approved to provide an effective classification tool for integrable quad-graph equations.

\section*{Acknowledgments}
The authors  thank A.V.Zhiber and R.I.Yamilov for valuable discussions. This work is partially supported by Russian Foundation for Basic
Research (RFBR) grants $\#$ 10-01-91222-CT-a, $\#$
08-01-00440-a, and $\#$ 10-01-00088-a.

\end{document}